\documentclass[11pt]{article}
\usepackage{acl2014}
\usepackage{times}
\usepackage{url}
\usepackage{latexsym}
\usepackage{graphicx}
\usepackage{comment}

\title{Predicting Central Topics in a Blog Corpus from a Networks Perspective}

\author{Srayan Datta \\
  University of Michigan \\
  Ann Arbor, Michigan 48105 \\
  {\tt srayand@umich.edu}
 }

\date{}

\begin{document}
\maketitle
\begin{abstract}
 In today's content-centric Internet, blogs are becoming increasingly popular and important from a data analysis perspective. According to Wikipedia, there were over 156 million public blogs on the Internet as of February 2011. Blogs are a reflection of our contemporary society. The contents of different blog posts are important from social, psychological, economical and political perspectives. Discovery of important topics in the blogosphere is an area which still needs much exploring. We try to come up with a procedure using probabilistic topic modeling and network centrality measures which identifies the central topics in a blog corpus.
\end{abstract}

\section{Introduction}
\label{sec:intro}

This paper presents an algorithm to identify and rank the most important topics given a set of blog entries. The topics are identified using Latent Dirichlet Allocation (LDA)~\cite{bleilda} -- a probabilistic topic model, and subsequently labeled to indicate their identity. Topic networks are created from the topics extracted using two different metrics. Graph centrality measures are run on these networks to find out two different rankings. A topic navigator is built to display the most relevant documents given a topic, where the topics are ranked by their importance given a similarity metric.

\section{Design}
\label{sec:design}

This section is divided into several parts. The first sub-section describes the dataset. The next one discusses data preprocessing and topic extraction, which is followed by the subsections topic network creation, centrality calculation, manual labeling and topic navigator description. 

\subsection{Dataset description}
\label{subsec:dataset_description}

Due to crawling being extremely time-consuming, a suitable publicly available blog corpus called the Blog Authorship Corpus ~\cite{DBLP:conf/aaaiss/SchlerKAP06}, is used for this project. This corpus contains around 681,000 posts from blogger.com. All of these blogs contain authorship information (to filter out some of the junk posts) and there are at least 200 words in each blog. The blog corpus contains 19,320 XML files where each file stores all the blog posts of a specific author.

\subsection{Data preprocessing and topic extraction}
\label{subsec:data_preprocessing}

The topic extraction is performed using LDA topic model, a very popular choice among different probabilistic topic models. The topic extraction is performed using a popular topic modeling toolkit called Mallet~\cite{McCallumMALLET}. As Mallet does not take XML files as input, each XML file is converted to a text file (.txt format) by stripping the XML metadata. Presence of stop words significantly deteriorates topic model performance, so stop word removal is done using Mallet.

Small text volume in a document affects LDA performance. Stop words represent a significant volume of any text. After stop word removal, the number of words in a blog entry decreases even more. All the blog entries by a single author are stored into a single file to alleviate this problem.  

Finding optimal number of topics given a corpus is a difficult problem. For the purpose of this project this number is assumed to be 100. Mallet outputs the most important 19 words for each topic and the probability distribution of the 100 topics over each file.  

\subsection{Manual labeling}
\label{subsec:manual_labeling}

As Mallet does not output the names of the topics, the top 19 words for each topic emitted by Mallet are used to manually label the topics. This is done so that each topic has an identity and is not just a bag of words. Assigning a name from just a list of words is a challenging task. In this case, this is even more challenging, as in the blog corpus, the language is very informal and there are languages other than English. There are cases where some proper nouns (names of persons) are returned as the most important words. These names do not give any idea about the underlying topic. Mallet is seen to output some mixed topics also. There are two topics for which no suitable nomenclature is found. These two topics are denoted by `Mixed'.      

\subsection{Topic network creation}
\label{subsec:topic_network_creation}

In this section, the methods of building topic networks from Mallet are discussed. A topic network is a weighted graph where each node is specified by a topic and an weighted edge between two nodes denotes the degree of similarity between the topics of those two nodes. Note that an edge weight is always positive. 

Building a topic network can be divided into two parts; identifying features for each topic and deciding which similarity measure works best. For the first part, we have two different feature vectors for each topic. Mallet outputs the probability of a word given a topic for each topic and each word (henceforth referred to as \emph{topic-word features}); and probability of a topic given a document for each document and each topic (henceforth referred to as \emph{topic-document features}). The topic-word feature becomes intractable due to a very high number of unique words in the dictionary. This number is very high as all derivatives of a single word are considered unique and there are words from languages other than English in the blog corpus. So, the topic-document features are used instead. Each topic $t_i$ is represented by a feature vector $<p_{ij}>$ of length 19,320 (number of documents in the blog corpus), where $p_{ij}$ is the probability that the $j^{th}$ document $d_j$ belongs to the $i^{th}$ topic $t_i$. These feature vectors are then used to calculate similarity between two topics.

Due to a lack of previous work of this nature, it is difficult to tell which similarity measure between the feature vectors will work the best in this case. So, instead of choosing a single measure, two very popular similarity measures were used to create two topic networks. The first one is cosine similarity~\cite{wiki:cosine_similarity}, which is a widely used metric in information retrieval and natural language processing, where it is used to measure similarity between documents, words, sentences, and so on. The other metric used is Pearson correlation coefficient~\cite{wiki:correlation} which is a measure of linear dependence between two random variables in statistics. 
 
Cosine similarity always lies between zero and one. So, when using this measure to build a topic network, there is always an edge with positive weight between any two nodes. So, the graph generated is a complete graph. On the other hand, correlation coefficient can take positive, negative and zero values. The negative and zero weight edges are pruned due to the fact that there is no well-known centrality measure which works well on a weighted graph with negative edge weights. So, the graph generated using correlation coefficient as a similarity measure is not a complete graph.

\subsection{Centrality calculation}
\label{subsec:centrality_calculation}

There are quite a few ways of assigning importance to nodes in a weighted graph using only graph structure. In this case we use the most famous one, Pagerank~\cite{Brin:1998:ALH:297810.297827}, which is shown to have good performance over a range of different networks including Internet link graph, citation graphs, social networks and word networks~\cite{mihalcea-tarau:2004:EMNLP}. Pagerank is run on both topic networks and the topics are ranked according to their Pagerank values.

Other popular centrality measures like degree and betweenness does not perform well because one of the topic networks is a complete graph. 

\subsection{Topic navigator description}
\label{subsec:topic_navigator_description}

A topic navigator augments the work presented above. The navigator provides a simple graphical user interface which allows a user to select a similarity metric from a drop-down list. Another drop-down menu showing the ranked list of topics according to the selected similarity measure is shown. This list is scrollable and selecting a topic shows the top 100 blogs related to the topic, ranked from most relevant to least relevant. Selecting a blog allows scrolling; double-clicking opens the document in the default text editor. Re-selecting a metric and a topic dynamically changes the results shown.

\begin{figure}[htb!]
 \centering
 \includegraphics[width=.5\textwidth]{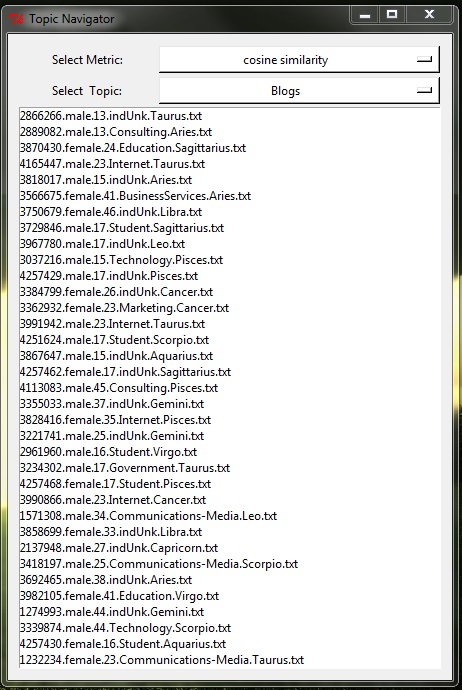}
 \caption{\label{Fig:navigator}The Topic Navigator}
\end{figure}

Figure \ref{Fig:navigator} shows a screen-shot of the topic navigator.

\section{Implementation and tools used}
\label{sec:implementation_tools_used}

The Topic Modelling is done using Mallet~\cite{McCallumMALLET}, as mentioned before. The rest of the implementation is done using python. Igraph library~\cite{igraph_cite} for python is used for pagerank implementation. The graph visualizations are generated using igraph and Gephi~\cite{ICWSM09154}. 

\section{Results}
\label{sec:results}

\begin{figure}[htb!]
 \centering
 \includegraphics[width=.3\textwidth]{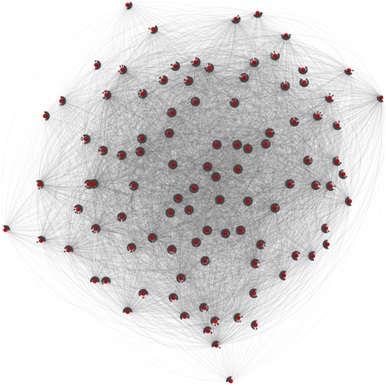}
 \caption{\label{Fig:cosine_network}The Topic Network using cosine similarity}
\end{figure}

\begin{figure}[htb!]
 \centering
 \includegraphics[width=.3\textwidth]{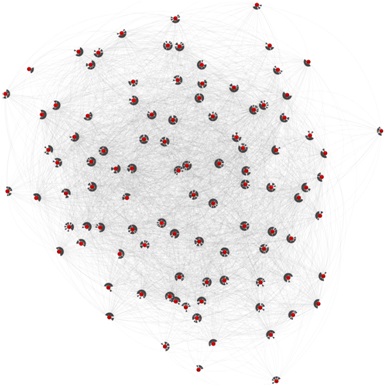}
 \caption{\label{Fig:corr_network}The Topic Network using correlation coefficient}
\end{figure}

This section presents the results obtained from the experiments described above. Figure \ref{Fig:cosine_network} presents a visualization of the topic network created using cosine similarity and figure \ref{Fig:corr_network} presents visualization of the topic network using correlation coefficient. These graphs give a broad idea about the nature of connection between the topics and the difference between the two topic networks.

We present top 10 topics determined by each similarity metric in a tabular format. The top 19 words for each topic are also presented. Table \ref{table:cosine} shows the top topics for cosine similarity measure and Table \ref{table:corr} likewise shows the same for correlation coefficient. Note that, some topics occur in both tables.

\begin{table}[htb!] 
\centering
\begin{tabular}{|p{0.3 in}|p{0.5 in}|p{1.8in}|} 
\hline 
Rank & Topic name & Top words \\
\hline
1 & Idle chitchat & ve ll don didn pretty time thing back things good lot haven doesn wasn bit stuff guess isn won  \\
\hline
2 & Spent time & time fact year years point day long days made bit apparently half completely spent entire thing end case simply  \\
\hline
3 & Talking about people & don people didn doesn thing things isn make good bad thought wasn feel hate won wouldn person talking talk  \\
\hline
4 & Good times & good time great cool pretty fun day today stuff hope guys thing guess lot nice back work things people  \\
\hline
5 & Swear words normal & yeah hell damn ll crap thing hey gonna blah ass stuff kinda back god guy stupid man gotta head  \\
\hline
6 & Work life & people time things make work good person feel life find don lot point part thing place change interesting mind  \\
\hline
7 & Time description & today work day night time home morning back tomorrow good tonight week sleep bed days yesterday hours house weekend  \\
\hline
8 & Fashion trends & hair make big day wear wearing favorite shoes white hot black shirt put pants clothes red things buy kind  \\
\hline
9 & Life and death & people man world evil death kill men dead die make great life human hell thing years war fight live \\
\hline
10 & How to write & ve words don word sense kind doesn real read high write sort writing isn makes line form piece sound \\
\hline
\end{tabular}
\caption{\label{table:cosine}Top topics determined from cosine similarity topic network}
\end{table}

\begin{table}[htb!] 
\centering
\begin{tabular}{|p{0.3 in}|p{0.5 in}|p{1.8in}|} 
\hline 
Rank & Topic name & Top words \\
\hline
1 & SMS lingo 1 & lol cuz gonna ya yea kinda wanna ppl omg today tho hey lil soo dunno stuff fun guys guess\\
\hline
2 & German words & tyke ich und tina die der ist das phil cuz mit nicht purdy dr mich es evan ein von\\
\hline
3 & SMS lingo 2 & da ur dat sum wat coz jus im bout wit ppl goin day hav skool wen lol rite nite \\
\hline
4 & Life and death & people man world evil death kill men dead die make great life human hell thing years war fight live \\
\hline
5 & Sexual preference & women sex gay men woman marriage sexual man male relationship married female love life lesbian straight make partner support \\
\hline
6 & Friends & alex jeremy mark hannah blog luke cindy day level fun time people lesson sammy school ted house cool love\\
\hline
7 & Education & school class students college law student classes year semester summer university campus teaching paper study professor high graduate teacher\\
\hline
8 & Today & alot day today ya thing bored back till house good time school bye kinda friends post life start friend\\
\hline
9 & Good times & good time great cool pretty fun day today stuff hope guys thing guess lot nice back work things people  \\
\hline
10 & Talking about people & don people didn doesn thing things isn make good bad thought wasn feel hate won wouldn person talking talk  \\
\hline
\end{tabular}
\caption{\label{table:corr}Top topics determined from correlation coefficient topic network}
\end{table}

\begin{figure}[htb!]
 \centering
 \includegraphics[width=0.5\textwidth]{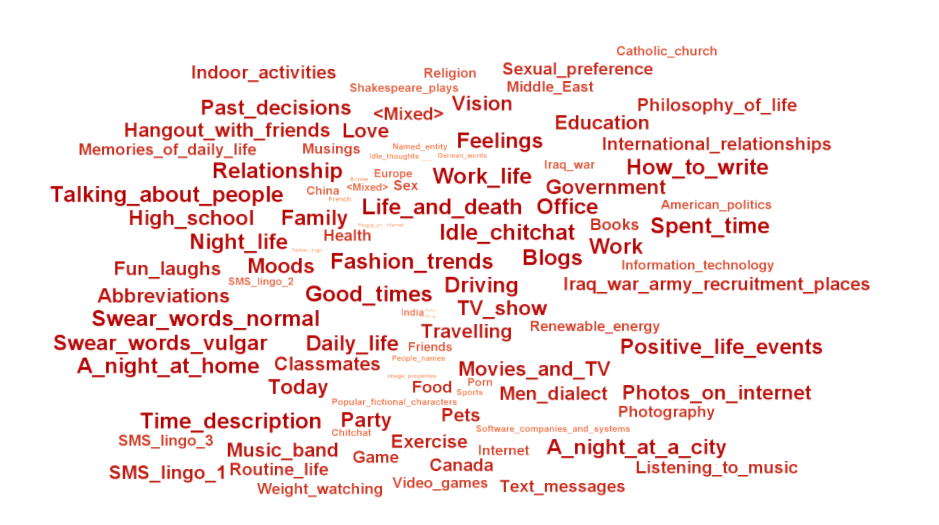}
 \caption{\label{Fig:cosine_cloud}The Topic Cloud using cosine similarity}
\end{figure}

\begin{figure}[htb!]
 \centering
 \includegraphics[width=0.5\textwidth]{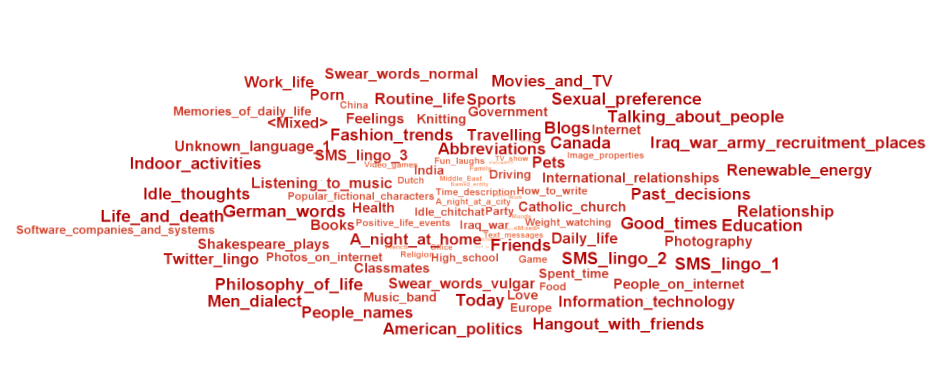}
 \caption{\label{Fig:corr_cloud}The Topic Cloud using correlation coefficient}
\end{figure}

We present visualizations of the ranked topics in the form of `topic cloud's . A topic cloud is a visual representation of topics in a corpus where the font size and color of the topic are determined by its importance. The larger and darker the topic, the more its importance. Note that unlike tag clouds (generally used for free-from text visualization), topic clouds maintain their underlying network structure. So, if two topics are nearby, they are more closely related to each other than two topics that lie at the two extreme ends of the picture. Figure \ref{Fig:cosine_cloud} shows the topic cloud derived from cosine similarity topic network. Likewise, figure \ref{Fig:corr_cloud} represents the topic cloud using correlation coefficient.

From the cosine similarity table, it is interesting to observe that the topics like `Idle chitchat' and `Talking about people' are ranked very high. Moreover, `Talking about people' is ranked high in correlation coefficient table. This, and the use of informal language, shows that a high number of blogs are used very casually in a conversational style. The top three topics in the correlation coefficient table clearly show that the top-ranked topics are influenced by the use of informal language and presence of foreign languages. Some potential improvements over these issues are discussed in the discussion and future works section.

This study is the first of its kind. So, there is no parallel set of results on topic ranking to compare with.

\section{Discussion and Future Work}
\label{sec:future_work}
Working with a blog corpus necessitates working with informal language, lots of abbreviations and wrong spellings. This makes topic extraction results look very different than the results reported in the literature (e.g., Table 2 of~\cite{Wang:2007:TNP:1441428.1442141}). Further, the blog corpus used in this project contains languages other than English. This makes those other languages (e.g., German) themselves emerge as `topics', rather than the actual topics that they encode. (A blog written in a different language outputs the language itself as its main topic irrespective of the content.) This happens because LDA assumes a topic is simply a probability distribution over words. Creating a special stop word list for the blog corpus and restricting the language to English may alleviate these problems.

Manual labeling from single words is difficult. Use of n-gram topic model~\cite{Wallach05topicmodeling,Wang:2007:TNP:1441428.1442141} can help with this issue, as it outputs phrases along with words, thereby potentially being more effective from a `topic labeling' viewpoint.

\section{Conclusion}
\label{sec:conclusion}

This study is the first of its kind. Topic modelling outputs the topics of a given corpus, the methodology described here provides a way to rank them. This paper describes how to extract important topics from blogs, but the algorithm provided is not limited to a blog corpus, it is domain and language independent. This model can be used to extract important topics from any monolingual corpus.

\section*{Acknowledgements}
\label{acknowledgements}
I would like to thank Professor Sugih Jamin for the topic navigator idea, and Mr. Shibamouli Lahiri for discussions about the project idea and help with manual labeling of topics.


\bibliographystyle{acl}
\bibliography{forpaper}

\begin{thebibliography}{}

\bibitem[\protect\citename{Bastian \bgroup et al.\egroup }2009]{ICWSM09154}
Mathieu Bastian, Sebastien Heymann, and Mathieu Jacomy.
\newblock 2009.
\newblock {Gephi: An Open Source Software for Exploring and Manipulating
  Networks}.

\bibitem[\protect\citename{Blei \bgroup et al.\egroup }2003]{bleilda}
David~M. Blei, Andrew~Y. Ng, and Michael~I. Jordan.
\newblock 2003.
\newblock {Latent Dirichlet Allocation}.
\newblock {\em Journal of Machine Learning Research}, 3:993--1022.

\bibitem[\protect\citename{Brin and Page}1998]{Brin:1998:ALH:297810.297827}
Sergey Brin and Lawrence Page.
\newblock 1998.
\newblock {The Anatomy of a Large-scale Hypertextual Web Search Engine}.
\newblock {\em Comput. Netw. ISDN Syst.}, 30(1-7):107--117, April.

\bibitem[\protect\citename{Csardi and Nepusz}2006]{igraph_cite}
Gabor Csardi and Tamas Nepusz.
\newblock 2006.
\newblock {The igraph software package for complex network research}.
\newblock {\em InterJournal}, Complex Systems:1695.

\bibitem[\protect\citename{McCallum}2002]{McCallumMALLET}
Andrew~Kachites McCallum.
\newblock 2002.
\newblock {MALLET: A Machine Learning for Language Toolkit}.
\newblock http://www.cs.umass.edu/~mccallum/mallet.

\bibitem[\protect\citename{Mihalcea and Tarau}2004]{mihalcea-tarau:2004:EMNLP}
Rada Mihalcea and Paul Tarau.
\newblock 2004.
\newblock {TextRank: Bringing Order into Texts}.
\newblock In Dekang Lin and Dekai Wu, editors, {\em Proceedings of EMNLP 2004},
  pages 404--411, Barcelona, Spain, July. Association for Computational
  Linguistics.

\bibitem[\protect\citename{Schler \bgroup et al.\egroup
  }2006]{DBLP:conf/aaaiss/SchlerKAP06}
Jonathan Schler, Moshe Koppel, Shlomo Argamon, and James~W. Pennebaker.
\newblock 2006.
\newblock {Effects of Age and Gender on Blogging}.
\newblock In {\em AAAI Spring Symposium: Computational Approaches to Analyzing
  Weblogs}, pages 199--205.

\bibitem[\protect\citename{Wallach}2005]{Wallach05topicmodeling}
Hanna~M. Wallach.
\newblock 2005.
\newblock Topic modeling: beyond bag-of-words.
\newblock In {\em NIPS 2005 Workshop on Bayesian Methods for Natural Language
  Processing}.

\bibitem[\protect\citename{Wang \bgroup et al.\egroup
  }2007]{Wang:2007:TNP:1441428.1442141}
Xuerui Wang, Andrew McCallum, and Xing Wei.
\newblock 2007.
\newblock Topical n-grams: Phrase and topic discovery, with an application to
  information retrieval.
\newblock In {\em Proceedings of the 2007 Seventh IEEE International Conference
  on Data Mining}, ICDM '07, pages 697--702, Washington, DC, USA. IEEE Computer
  Society.

\bibitem[\protect\citename{Wikipedia}2013a]{wiki:cosine_similarity}
Wikipedia.
\newblock 2013a.
\newblock Cosine similarity --- {W}ikipedia{,} the free encyclopedia.
\newblock [Online; accessed 28-November-2013].

\bibitem[\protect\citename{Wikipedia}2013b]{wiki:correlation}
Wikipedia.
\newblock 2013b.
\newblock Pearson product-moment correlation coefficient --- {W}ikipedia{,} the
  free encyclopedia.
\newblock [Online; accessed 28-November-2013].

\end{thebibliography}

\end{document}